\documentclass[10pt, prl,aps,twocolumn,nofootinbib]{revtex4-1}
\usepackage{amssymb,amsmath,amsfonts,amsbsy,epsfig,graphicx}
\usepackage[dvipsnames]{xcolor}
\usepackage{braket}
\definecolor{lgray}{gray}{0.35}
\usepackage[colorlinks=true,urlcolor=Gray,linkcolor=lgray,
citecolor=Gray,
             pdfpagelabels=true,hypertexnames=true,
            plainpages=false,naturalnames=false,
             ]{hyperref}

\newcommand{\be}{\begin{equation}}
\newcommand{\ee}{\end{equation}}
\newcommand{\bea}{\begin{eqnarray}}
\newcommand{\eea}{\end{eqnarray}}
\newcommand{\nn}{\nonumber}

\begin{document}

\title{Reconstructing the Inflationary Landscape with Cosmological Data}

\date{October 17, 2018}

\author{
Xingang Chen$^{a}$, Gonzalo A. Palma$^{b}$, Bruno Scheihing H.$^{b}$ and Spyros Sypsas$^{b}$
}

\affiliation{
$^{a}$Institute for Theory and Computation, Harvard-Smithsonian Center for Astrophysics, 60 Garden Street, Cambridge, MA 02138, USA. \\
$^{b}$Grupo de Cosmolog\'ia y Astrof\'isica Te\'orica, Departamento de F\'{i}sica, FCFM, \mbox{Universidad de Chile}, Blanco Encalada 2008, Santiago, Chile.}

\begin{abstract}

We show that the shape of the inflationary landscape potential may be constrained by analyzing cosmological data. The quantum fluctuations of fields orthogonal to the inflationary trajectory may have probed the structure of the local landscape potential, inducing non-Gaussianity (NG) in the primordial distribution of the curvature perturbations responsible for the cosmic microwave background (CMB) anisotropies and our Universe's large-scale structure. The resulting type of NG (tomographic NG) is determined by the shape of the landscape potential, and it cannot be fully characterized by 3- or 4-point correlation functions. Here we deduce an expression for the profile of this probability distribution function in terms of the landscape potential, and we show how this can be inverted in order to reconstruct the potential with the help of CMB observations. While current observations do not allow us to infer a significant level of tomographic NG, future surveys may improve the possibility of constraining this class of primordial signatures.

\end{abstract}

\maketitle

Is there any feature about our Universe that would require us to assume primordial non-Gaussian initial conditions? Up until now, cosmic microwave background (CMB) and large-scale structure (LSS) observations are fully consistent with the premise that the primordial curvature perturbations were initially distributed according to a perfectly Gaussian statistics~\cite{Ade:2015ava, Komatsu:2003fd}. This has favored the simplest models of inflation --single field slow-roll inflation-- based on the steady evolution of a scalar field driven by a flat potential~\cite{Guth:1980zm, Starobinsky:1980te, Linde:1981mu, Albrecht:1982wi, Mukhanov:1981xt}. In these models, the self-interactions of the primordial curvature perturbation lead to tiny non-Gaussianities suppressed by the slow-roll parameters characterizing the evolution of the Hubble expansion rate $H$, during inflation~\cite{Gangui:1993tt, Komatsu:2001rj, Acquaviva:2002ud, Maldacena:2002vr}.

The confirmation of non-Gaussian initial conditions would help us to decipher certain fundamental aspects about inflation~\cite{Bartolo:2004if, Liguori:2010hx, Chen:2010xka, Wang:2013eqj}. Indeed, non-Gaussianity (NG) can be generated by nonlinearities affecting the evolution of primordial curvature perturbations (denoted as $\zeta$). These nonlinearities are the result of self-interactions, or interactions with other degrees of freedom, such as isocurvature fields (fields orthogonal to the inflationary trajectory in multifield space). Inevitably, perturbation theory limits the extent to which we can study the emergence of NG, forcing us to focus on the lowest order operators (in terms of field powers) in the $\zeta$ Lagrangian. Thus, most of the recent effort devoted to the study of NG has relied on parametrizing it with the bispectrum and trispectrum, the amplitudes of the $3$- and $4$-point correlation functions of $\zeta$ in momentum space. Understanding how different interactions lead to different shapes and runnings of the bispectrum has constituted one of the main programs in the study of inflation~\cite{Bartolo:2004if, Liguori:2010hx, Chen:2010xka, Wang:2013eqj}.

It is conceivable, however, that certain classes of interactions may lead to NG deviations that cannot be parametrized just with the bispectrum and/or trispectrum. This is the subject of the companion article~\cite{Chen:2018uul}, where we argue that in multifield models characterized by a rich landscape structure (i.e., with minima separated by field distances of order, or smaller than, $H$), extra fields can transfer their NG to $\zeta$. In two-field models, the mechanism by which this NG is generated relies on the existence of an isocurvature field $\psi$ interacting with $\zeta$ via a generic coupling that appears in multifield models. The mechanism may be understood as the consequence of the following two independent statements:
\begin{enumerate}

 \item[I.] If on superhorizon scales the amplitude of $\psi$ does not vanish, then it will act as a source for the amplitude of $\zeta$. The field $\zeta$ will grow on superhorizon scales and become related to $\psi$ (e.g.,~\cite{Gordon:2000hv,Chen:2009zp,Achucarro:2016fby}).

\item[II.] If $\psi$ has a potential $\Delta V (\psi)$ with a rich structure, then around horizon crossing $\psi$ will fluctuate and diffuse across the potential barriers. After horizon crossing, it will be more probable to measure $\psi$ at values that minimize $\Delta V (\psi)$~\cite{Palma:2017lww}.
\end{enumerate}
Together, these two statements imply that the probability of measuring $\zeta$ is higher at those values sourced by $\psi$ that minimize $\Delta V$. This was shown in~\cite{Chen:2018uul} for the particular case in which $\psi$ is an axionlike field, with $\Delta V = \Lambda^4 \left[ 1 - \cos (\psi / f) \right]$. There, the main result consisted in the derivation of a probability distribution function (PDF) $\rho (\zeta)$ that depended explicitly on the barrier height $\Lambda^4$ and the field range $f$.

The purpose of this Letter is to extend the derivation of~\cite{Chen:2018uul} to an arbitrary analytic potential $\Delta V$, and to show how it is possible to reconstruct its shape with current and/or future cosmological data. Our main claim is that, if the primordial landscape had a rich structure, then its shape (around the inflationary trajectory) could be stored in the statistics of $\zeta$ through a type of NG (tomographic NG) that cannot be fully parametrized with the bispectrum alone.

Our starting point is to consider the following generic Lagrangian describing $\zeta$ and $\psi$ ($M_{\rm Pl} = 1$):
\be
\mathcal L = a^3 \Big[\epsilon  ( \dot \zeta - \alpha \psi ) ^2 -   \frac{\epsilon}{a^2} (\nabla \zeta)^2  +  \frac{1}{2} \dot \psi^2  - \frac{1}{2a^2} (\nabla \psi)^2   -  \Delta V \Big], \label{Lagrantian-full-v}
\ee
where $a$ is the scale factor, and $\epsilon = - \dot H /H^2$ is the usual first slow-roll parameter ($H = \dot a/a$). In this system, $\zeta$ interacts with $\psi$ via $\mathcal L_{\rm int} \propto \alpha \dot \zeta \psi$ (with $\alpha$ constant). Note that we are treating both $\zeta$ and $\psi$ up to quadratic order, except for $\psi$ appearing in $\Delta V$. We assume that $\Delta V /V_{\rm infl} \ll 1$ so that inflation, driven by $V_{\rm infl}=3M_{\rm Pl}^2H^2$, is unaffected by $\Delta V$. In what follows, ${\bf x}$ and ${\bf k}$ denote comoving position and momentum, whereas ${\bf q} = {\bf k} /a$ denotes physical momentum.

If $\Delta V = 0$, Eq.~(\ref{Lagrantian-full-v}) gives us two linear equations of motion for $\zeta$ and $\psi$ coupled through $\alpha$. In $k$ space, the dynamics is such that the mode function $\psi_{k} (t)$ becomes frozen to a constant value at horizon crossing. Then, $\psi_{k}$ acts as a source for the amplitude of $\zeta_{k}$, and one finds~\cite{Achucarro:2016fby}
\be
\zeta_{k} = (\alpha \Delta N / H) \psi_{k} , \label{source-zeta}
\ee
where $\Delta N$ is the number of $e$-folds after horizon crossing. As a result, the power spectrum of $\zeta$ is determined by that of $\psi$ as $P_\zeta (k) = \frac{\alpha^2 \Delta N^2}{H^2} P_{\psi} (k)$. Thus, the field $\psi$ transfers its Gaussian statistics to $\zeta$ via $\alpha$.

On the other hand, if $\Delta V \neq 0$, the field $\psi$ continues to transfer its statistics to $\zeta$ (thanks to $\alpha$), but this time it will inherit NG deviations. In momentum space, $\Delta V$ induces nonvanishing $n$-point correlation functions of the local type, given by
\be \label{n-point-local}
\langle \zeta_{{\bf k}_1 ... {\bf k}_n}^n \rangle_c =  (2\pi)^3 \, h_n \, \delta^{(3)} \Big( \sum_{i=1}^n {\bf k}_i \Big) \frac{k_1^3 + \cdots + k_n^3}{k_1^3 \cdots k_n^3} ,
\ee
where $c$ informs us that we are only keeping fully connected contributions (in the language of perturbation theory). To obtain the set of amplitudes $h_n$ for an arbitrary potential we first consider the following Taylor expansion
\be \label{potential-taylor}
\Delta V \left( \psi \right) = \sum_{m} \frac{c_m}{m!} \psi^m.
\ee
This expansion gives us an infinite number of $m$ legged vertices, each one of order $c_m$.  Using the in-in formalism, the computation of $\langle \zeta_{{\bf k}_1 ... {\bf k}_n}^n \rangle_c$ requires us to consider the sum of each Feynman diagram proportional to $c_{n + 2m}$ with $m \geq 0$. In any such diagram, $n$ legs become $\zeta$ external legs (due to the $\alpha$ coupling), whereas $2m$ legs become $m$ loops. Finally, $\langle \zeta_{{\bf k}_1 ... {\bf k}_n}^n \rangle_c$ is the result of summing all of these diagrams after taking into account the appropriate combinatorial factors. One finds
\be
h_n =-\left( \frac{\alpha H \Delta N}{2 } \right)^n \frac{\Delta N}{3H^{4}} \sum_{m=0}^{\infty} \frac{c_{n+2m}}{m!}  \left(\frac{\sigma_0^2}{2}  \right)^{m} , \label{h-n-V}
\ee
where $\sigma_0^2  \equiv  (2\pi)^{-3} \! \int d^3 k \, \psi_k^*(t) \psi_k(t)$, appearing because of the loops, is the variance of the field $\psi$.
Here, $\psi_k(t)$ is the mode function of a free massless field in a de Sitter spacetime. It turns out that $\sigma_0^2$ is time independent~\cite{Palma:2017lww}. 

Performing the sum in Eq.~(\ref{h-n-V}), one obtains
\be
\sum_{m=0}^{\infty} \frac{c_{n+2m}}{m!}  \left(\frac{\sigma_0^2}{2}  \right)^{m} = e^{\frac{\sigma_0^2}{2} \partial_\psi^2} \frac{\partial^n}{\partial \psi^n} \Delta V \bigg|_{\psi = 0} . \label{general-resummation}
\ee
Notice that $\sigma_0^2$ is formally infinite, and hence, we are forced to introduce infrared (IR) and ultraviolet (UV) physical momentum cutoffs. The UV cutoff $q_{\rm UV}$ corresponds to a wavelength well inside the horizon ($q_{\rm UV} \gg H$), whereas the IR cutoff $q_{\rm IR}$ corresponds to the wavelength of the largest observable mode. In addition to these scales, it is convenient to introduce an arbitrary intermediate momentum $q_L$ that splits $\sigma_0^2$ into two contributions: $\sigma_0^2 = \sigma_S^2 + \sigma_L^2$, from short and long modes, respectively. This splitting allows us to define a renormalized potential $\Delta V_{\text{ren}} (\psi) \equiv \exp \left( \frac{\sigma_S^2}{2} \frac{\partial^2}{\partial \psi^2} \right) \Delta V(\psi)$. In this way, observables can only depend on $\Delta V_{\text{ren}}$, which is independent of $q_{\rm UV}$.  

According to Eq.~(\ref{general-resummation}), this renormalization procedure simply corresponds to defining $\Delta V_{\rm ren} (\psi) = \sum_m c_m^{\rm ren} \psi^m / m!$, where the coefficients $c_m^{\rm ren}$ are related to the bare couplings $c_m$ as
\be
\sum_{m=0}^{\infty} \frac{c_{n+2m}}{m!}  \left(\frac{\sigma_0^2}{2}  \right)^m = \sum_{m=0}^{\infty} \frac{c_{n+2m}^{\text{ren}}}{m!}  \left(\frac{\sigma_L^2}{2}  \right)^m. \label{coeff-ren}
\ee
This result allows us to identify $\Delta V_{\text{ren}}$ as the potential obtained by integrating out the high energy momenta beyond the scale $q_L$, just as in the Wilsonian approach of QFT. Now, it is crucial to notice that the $n$-point function of Eq.~(\ref{n-point-local}) is an observable, and so it cannot depend on $q_{L}$. This implies that $h_n$ is independent of $\sigma_L$. For this to be possible, the coefficients $c_{m}^{\rm ren}$ defining $\Delta V_{\rm ren}$ must run in such a way so that the entire expression (\ref{h-n-V}) remains independent of $\sigma_L$. Equation~(\ref{coeff-ren}) reveals how the coefficients $c_m^{\rm ren}$ run as more (or fewer) modes participate in $\sigma_L^2$ (again, in agreement with the Wilsonian picture).

To continue, using the Weierstrass transformation, the right hand side of Eq.~(\ref{general-resummation}) can be rewritten as
\be
e^{\frac{\sigma_L^2}{2} \partial_\psi^2} \frac{\partial^n}{\partial \psi^n} \Delta V_{\rm ren} \bigg|_{\psi = 0} = \int \! d\psi \frac{e^{- \frac{\psi^2}{2 \sigma_L^2}} }{\sqrt{2 \pi} \sigma_L}  \frac{\partial^n}{\partial \psi^n}  \Delta V_{\rm ren} .
\ee
Then, by performing several partial integrations, we finally obtain the following expression for $h_n$:
\bea \label{h-n-V-2}
h_n &=&   \frac{1}{n}  \left( \frac{\alpha H \Delta N}{2 \sigma_L} \right)^n \frac{ \Delta N}{3H^{4}}  \int \!  d\psi \frac{e^{-\frac{\psi^2}{2\sigma_L^2}}}{\sqrt{2\pi} \sigma_L} {\rm He}_n\left( \psi /\sigma_L \right) \nn \\
&&
\times  \left( \sigma_L^2 \frac{\partial^2}{\partial \psi^2} - \psi \frac{\partial}{\partial \psi} \right) \Delta V_{\text{ren}} (\psi) , \quad
\eea
where ${\rm He}_n(x) \equiv \exp( - \frac{1}{2} \frac{d^2}{dx^2}) x^n$ is the $n$th ``probabilist's" Hermite polynomial. In the particular case where $\Delta V (\psi) = \Lambda^4 \left[ 1 - \cos (\psi / f) \right]$, Eq.~(\ref{h-n-V-2}) allows us to recover the expression for $\langle \zeta_{{\bf k}_1... {\bf k}_n}^n \rangle_c$ obtained in~\cite{Chen:2018uul}.

We now compute the $n$th moment $\langle \zeta^n \rangle$ for a particular position ${\bf x}$. Because of momentum conservation, the specific value of ${\bf x}$ is irrelevant. In practice, we only have observational access to a finite range of scales, implying that the computation of $\langle \zeta^n \rangle$ must consider a window function selecting that range. We use a window function with a hard cutoff, and write
\be
\zeta_L =  \frac{1}{(2 \pi)^3}  \int_{ k < k_{L}} \!\!\!\!\!\!\!\!\! d^3k \,  \zeta_{\bf k}  \, e^{- i {\bf k} \cdot {\bf x}} .
\ee
Notice that we have chosen to cut the integral with the same cutoff $k_{L} = a \, q_L$ introduced to split $\sigma_0^2 = \sigma_S^2 + \sigma_L^2$. Up until now, $q_L$ was an arbitrary scale introduced to select the scales integrated out to obtain $\Delta V_{\rm ren}$. However, we can now choose $q_L$ to coincide with the physical cutoff momentum setting the range of modes contributing to the computation of $\langle \zeta^n_L \rangle$. Given that we are interested in a $q_L^{-1}$ larger than the horizon, we can write
\be
\sigma_L^2 = ( H^2 / 4 \pi^2) \ln \xi , \label{sigma_L-res}
\ee
where $\xi \equiv k_L / k_{\rm IR}$. Following our companion paper~\cite{Chen:2018uul}, the $n$th moment of $\zeta_L$ is given by
\bea
\langle \zeta_L^n \rangle_c &=&  (2\pi)^3 \, h_n \, I_n (\xi) , \label{n-point-x-space}  \\
I_n (\xi) &=& \frac{n}{ (2 \pi^2)^{n+1}} \int_0^{\infty} \!\! \frac{dx}{x} G_\xi(x) \left[ F_\xi(x) \right]^{n-1} , \label{I_n-integral}
\eea
where $G_\xi(x) =  \int^1_{\xi^{-1}}   d z z x^2 \sin (z x)$ and $F_\xi(x)=\int^1_{\xi^{-1}}\frac{d y}{y} \frac{\sin (y x)}{y x} $. The function $F_\xi (x)$ satisfies $F_\xi(0) = \ln \xi$, and $F_\xi (x) \leq \ln \xi$. The PDF $\rho(\zeta)$ must be such that
\be
\langle \zeta_L^n \rangle = \int \! d \zeta \, \rho (\zeta) \zeta^n ,
\ee
where $\langle \zeta_L^n \rangle$ is the full $n$th moment, including disconnected contributions, related to $\langle \zeta_L^n \rangle_c$ by $\langle \zeta_L^n \rangle = \sum_{m=0}^{\lfloor n/2 \rfloor} \frac{n!}{m! (n-2m)! 2^m} \sigma_\zeta^{2m} \langle \zeta_L^{n-2m} \rangle_c$. Here, $\sigma_{\zeta}^2$ is the variance of $\zeta$, which according to Eqs.~(\ref{source-zeta}) and (\ref{sigma_L-res}), is given by $\sigma_{\zeta}^2 = \alpha^2 \Delta N^2 \sigma_{L}^2 / H^2 = \left[\alpha^2 \Delta N^2  / \left(4 \pi^2\right)\right] \ln \xi $. Planck fixes $\sigma_{\zeta}^2/ \ln \xi = P_{\zeta}(k) k^3 / \left(2 \pi^2\right) = (2.196 \pm 0.158 )\times 10^{-9}$.

To derive $\rho (\zeta)$ we just need to focus on the $n$ dependence of $\langle \zeta^n \rangle_c$. According to Eq.~(\ref{n-point-x-space}), this dependence has the form $[X]^{n-1} {\rm He}_n (Y)$, where $X$ and $Y$ are given quantities (the presence of the integrals do not alter this argument). This alone allows us to infer the PDF for $\zeta$, which is found to be given by
\bea
\rho (\zeta) &=&\frac{1}{\sqrt{2\pi}\sigma_\zeta } e^{-\frac{\zeta^2}{2\sigma_\zeta^2}}  \left[ 1 + \Delta(\zeta) \right],  \label{main-result-1} \\
 \Delta (\zeta) &\equiv& \int_0^{\infty} \!\! \frac{dx}{x}  \mathcal{K}(x) \! \int_{-\infty}^{\infty} \!\!\!\! d \bar \zeta \,\, \frac{\exp \Big[{-\frac{ \left(\bar \zeta - \zeta(x) \right)^2}{2\sigma_\zeta^2 (x)} } \Big] }{\sqrt{2\pi  } \sigma_\zeta (x)} \nn \\
 && \times \frac{\Delta N}{3 H^4}  \left( \sigma_\zeta^2 \frac{\partial^2}{\partial {\bar \zeta}^2} - {\bar \zeta} \frac{\partial}{\partial {\bar \zeta}} \right) \Delta V_{\text{ren}}  \! \left(\psi_{\bar \zeta}\right) . \label{main-result-2}
\eea
In the previous expression, $\Delta(\zeta)$ parametrizes the NG deviation. To write it, we defined the following quantities: $\zeta(x) \equiv   [ F_\xi(x) /\ln \xi ]  \zeta$, $\sigma_{\zeta}^2 (x) \equiv  \sigma_{\zeta}^2 (1 - [ F_\xi(x) /\ln \xi]^2 )$, $\mathcal{K}(x) \equiv 4\pi G_\xi(x) / F_\xi(x)$, and $\psi_{\zeta} \equiv (\alpha \Delta N/ H)^{-1} \zeta$. These definitions satisfy $|\zeta (x)| \leq |\zeta |$ and $\sigma_{\zeta}^2 (x) \leq \sigma_{\zeta}^2$.

\begin{figure}[t!]
\includegraphics[scale=0.65]{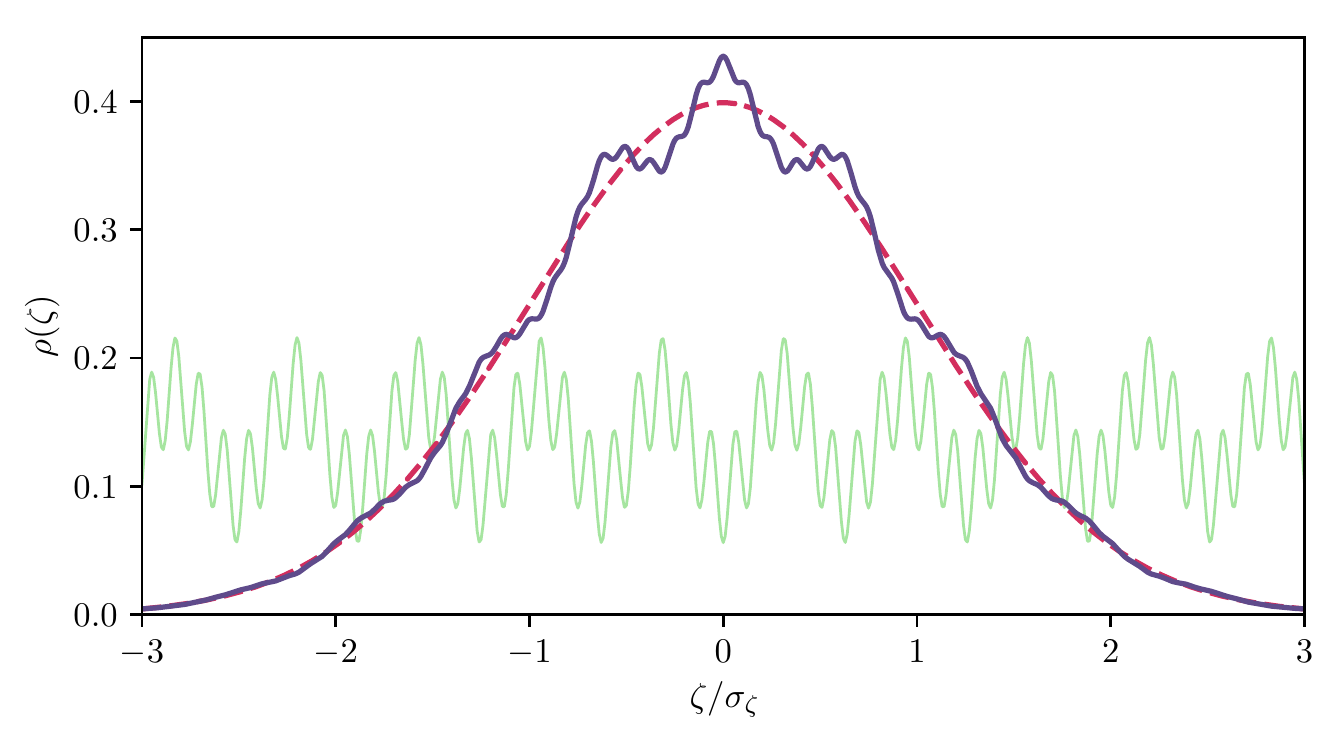}
\caption{The PDF (violet) resulting from a potential $\Delta V (\psi) \propto \left[ 2 - \cos (\psi/f_1) - \cos (\psi/f_2) \right]$, with $f_1 = 0.1 \sigma_L$ and $f_2 = 0.02 \sigma_L$ (light green). Both contributions have the same amplitude; however, $f_2$ contributes less than $f_1$. A Gaussian PDF is plotted for comparison (red, dashed).}
\label{fig:PDF_example}
\end{figure}

Equation~(\ref{main-result-1}) gives us the PDF of $\zeta$ at the end of inflation. It is possible to verify that the perturbativity condition on the potential is $\Delta V_{\rm ren} / H^4 \ll 1$, and that the next to leading order term is of order $\mathcal O (\Delta^2)$ (see Ref.~\cite{Chen:2018uul}). The presence of the derivative operator acting on $\Delta V_{\rm ren}$ implies that the probability of measuring $\zeta$ at a given amplitude increases at those corresponding values $\psi_{\zeta}$ that minimize the potential. In addition, the $x$ dependence of $\zeta(x)$ and $\sigma_{\zeta}^2 (x)$ has the effect of filtering the structure; sharper structures contribute less to the PDF. Figure~\ref{fig:PDF_example} shows the PDF obtained for a potential $\Delta V_{\rm ren} (\psi) \propto \left[ 2 - \cos (\psi/f_1) - \cos (\psi/f_2) \right]$. In this example there are two sinusoidal contributions with field scales $f_1 = 0.1 \sigma_L$ and $f_2 = 0.02 \sigma_L$. Both contributions have the same amplitude, however, the NG deformation implied by $f_2$ is smaller than that of $f_1$. Notice that to plot the figure, we used the relation $\psi_\zeta / \sigma_L = \zeta / \sigma_\zeta$.

Let us now attempt to reconstruct $\Delta V_{\rm ren}$ out of the CMB data. This requires us to deal with the observed temperature fluctuation $\Theta \equiv \Delta T / T$, instead of $\zeta$ at the end of inflation. Thus, we introduce a linear transfer function to write $\Theta ( {\bf k} , \hat n) \equiv T(k, \mu) \zeta_{\bf k}$, with $\mu = \hat n \cdot \hat k$, where $\hat n$ is the direction of sight of an observer standing at ${\bf x}$.  It follows that $\langle \Theta_{{\bf k}_1 ... {\bf k}_n}^n \rangle_c =   T(k_1 , \mu_1) \cdot \cdot \cdot T(k_n , \mu_n) \langle \zeta_{{\bf k}_1 ... {\bf k}_n}^n \rangle_c$, from which we are able to derive the connected $n$th moment:
\be
 \langle \Theta_L^n \rangle_c =  (2\pi)^3  h_n \left( \sigma_\Theta / \sigma_\zeta \right)^n  I^T_n (\xi ) , \label{n-point-temp-x-space}
\ee
where $I^T_n (\xi )$ is given by
\bea
I^T_n \! &=& \! \frac{n  }{2 (2 \pi^2)^{n+1}} \! \int_{-1}^{+1} \!\!\!\!\!\!\! d\mu \! \int_0^{\infty} \!\! \frac{dx}{x} G^T_\xi ( x , \mu) \! \left[ F^T_\xi( x , \mu) \right]^{n-1} \!\! , \,\,  \quad \label{I_n-integral-mu}
\eea
\bea
 G^T_\xi \! &=& \! \frac{\sigma_\zeta}{\sigma_\Theta} \sum_{\ell } (2 \ell + 1) P_\ell ( \mu ) \!\! \int_{\xi^{-1}}^1 \!\!\!\!\!\! d z z^2 x^3  T_\ell ( z k_L ) j_{\ell}(z x) , \qquad  \label{G-T} \\
 F^T_\xi \! &=& \! \frac{\sigma_\zeta}{\sigma_\Theta} \sum_{\ell } (2 \ell + 1)  P_\ell ( \mu ) \!\! \int_{\xi^{-1}}^1 \!\!\! \frac{d y }{y}  T_\ell ( y k_L ) j_{\ell}(y x) . \qquad  \label{F-T}
\eea
In the previous expressions, $P_\ell(x)$ and $j_\ell (x)$ stand for the $\ell$th Legendre polynomial and $\ell$th spherical Bessel function, respectively.  In addition, $T_\ell$ is the Legendre moment of $T(k, \mu )$. The variance of $\Theta$ is found to be $\sigma_\Theta^2 =  \frac{1}{4 \pi} \sum_{\ell }  (2\ell + 1) C_\ell $, with $C_\ell = 4 \pi [ \sigma_\zeta^2 / \ln \xi ] \int^{k_L}_{k_{\rm IR}} \frac{dk}{k}  |T_\ell (k)|^2$. 

One can now derive a PDF $\rho (\Theta)$ for $\Theta$ similar to that of Eqs.~(\ref{main-result-1}) and (\ref{main-result-2}). However, we will not need an explicit expression for $\rho (\Theta)$ to engage in reconstructing $\Delta V$. Instead, we may define the following cumulants parametrizing NG:
\be
a_n  \equiv    \int d\Theta \; \rho(\Theta) \, {\rm He}_n\left( \Theta / \sigma_\Theta \right)  . \label{pdf-a_n}
\ee
Independently of the form of $\rho(\Theta) $, these coefficients are directly related to the fully connected moments of $\Theta$ through the relation $\langle \Theta_L^n \rangle_c = \sigma_\Theta^n a_n$. Together with (\ref{n-point-temp-x-space}), this further implies
\be
h_n= a_n \sigma_\zeta^{n}  / [ (2\pi)^3  I^T_n (\xi )] . \label{a-n-h-n}
\ee
Then, by expanding the potential in terms of Hermite polynomials $\Delta V_{\text{ren}}(\psi)/H^4 = \sum_{m} \frac{b_m}{m!} \, {\rm He}_m \! \left( \psi / \sigma_L \right)$, one finds that the coefficients $b_n$ determining the shape of the potential are given by
\be
 b_n  = - \frac{3 a_n }{(2\pi)^3 \Delta N I^T_n(\xi)} \left( \frac {\ln \xi}{2 \pi^2}  \right)^n . \label{coeff-b_n}
\ee
The potential $\Delta V_{\text{ren}}(\psi)$ obtained by such a reconstruction has renormalized coefficients $c_{m}^{\rm ren}$ evaluated at the scale $k_{L}$, and so it can be interpreted as the potential generating NG in the range $k_{\rm IR} \leq k \leq k_L$.

\begin{figure}[t!]
\includegraphics[scale=0.58]{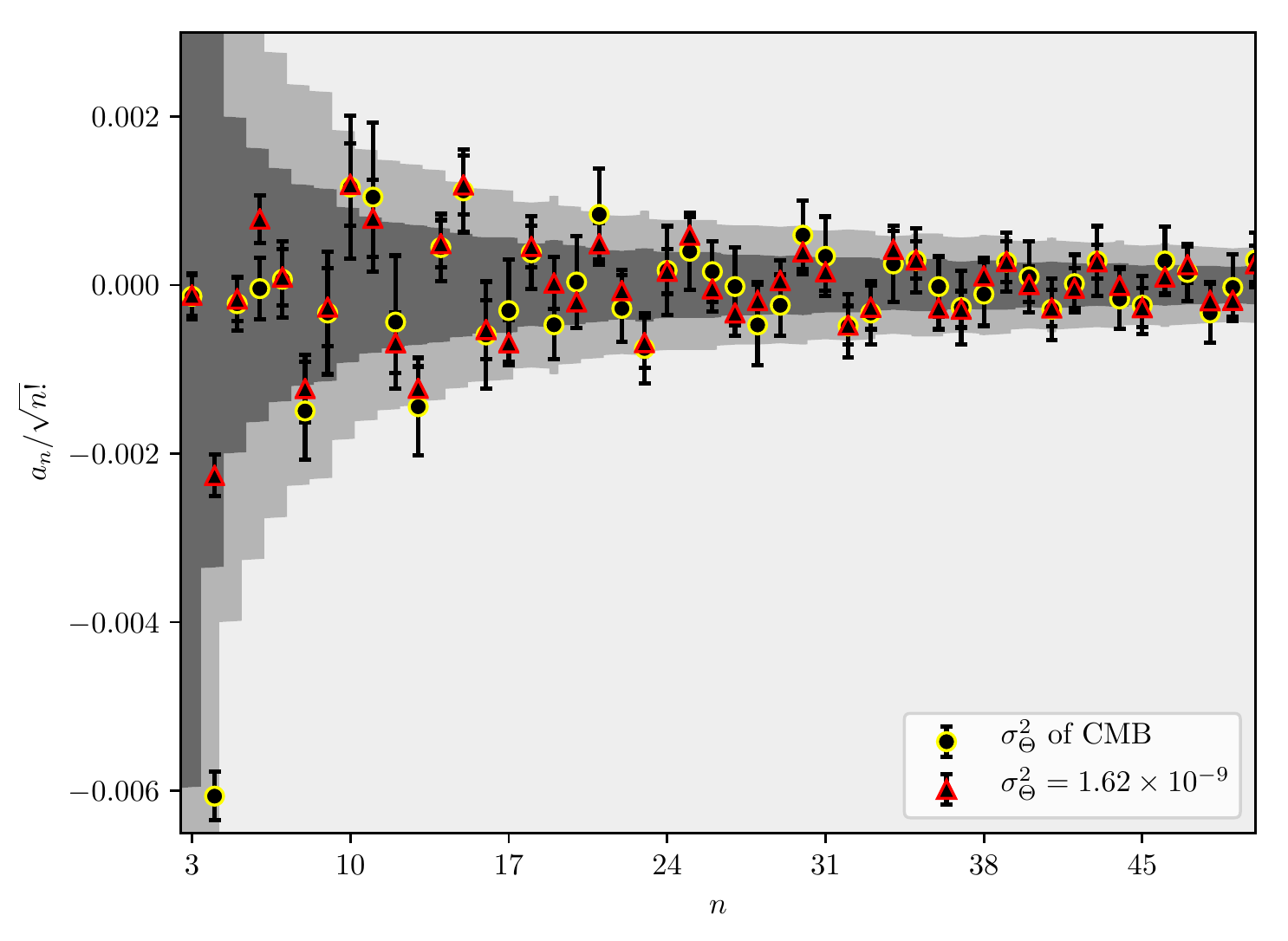}
\caption{The $a_n$ coefficients obtained from Planck. We have limited the data to regions far enough from the galactic plane so that the outcome from SMICA agrees with the other pipelines (and removing each pipeline's masked pixels), effectively considering a third of the sky. The error bars are an estimate of the noise present in the data, computed by comparing half-mission maps. The grey contours represent the intrinsic noise $\sigma(a_n)$ due to Gaussian simulations obtained using full-sky maps generated with CAMB.}
\label{fig:a_n-Planck}
\end{figure}

\begin{figure}[t!]
\includegraphics[scale=0.55]{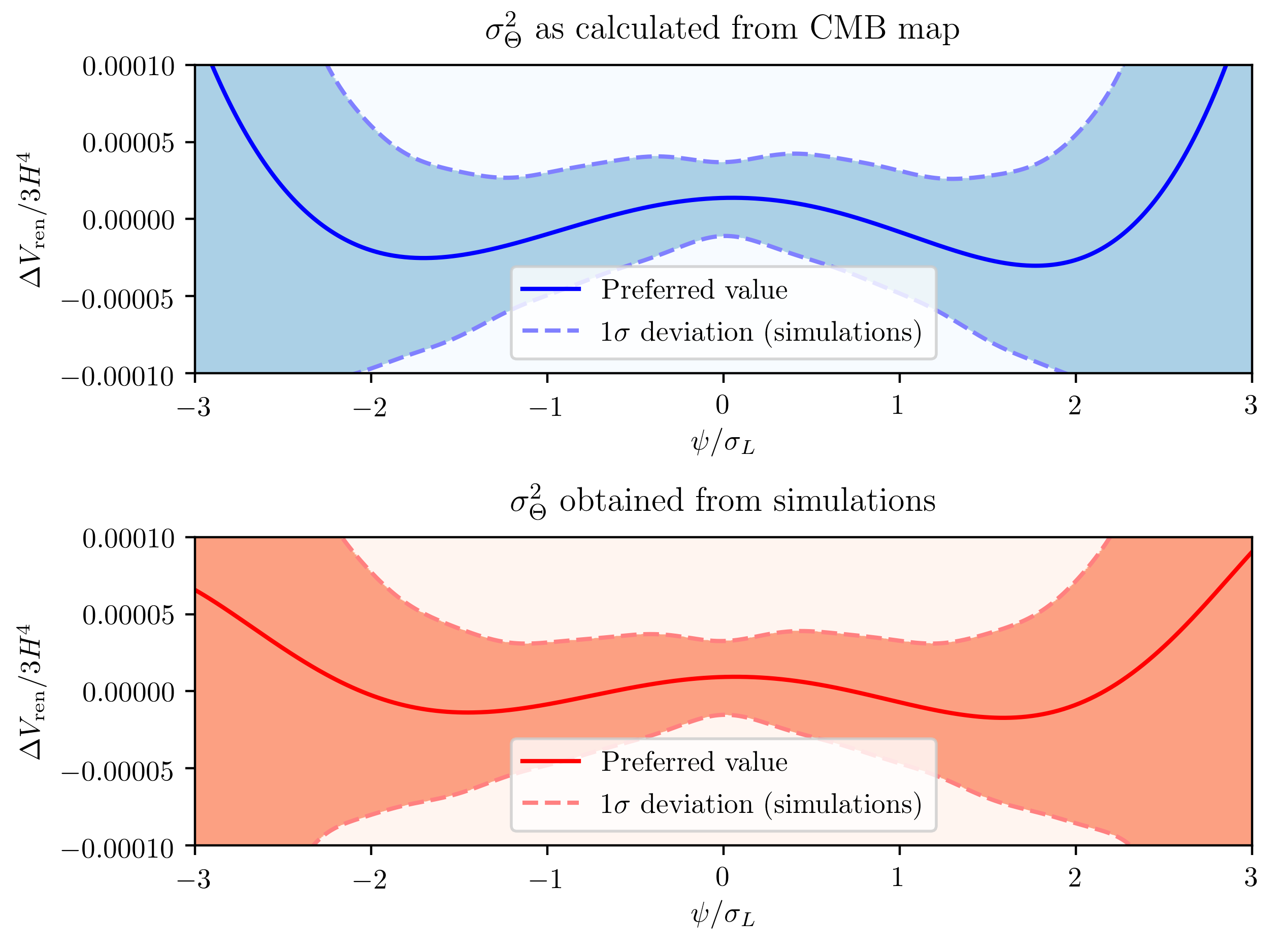}
\caption{The reconstructed potential $\Delta V/3 H^4$ for two different values of $\sigma_\Theta^2$. This reconstruction considers $a_n$ coefficients up to $n=7$. Since $a_2$ is a correction to the 2-point function --and hence, to the propagator-- we do not include this term in the reconstructed potential.}
\label{fig:DeltaV-reconstructed}
\end{figure}

Having Eq.~(\ref{coeff-b_n}) at hand, we may proceed to outline the reconstruction process. Figure~\ref{fig:a_n-Planck} shows values of the coefficients $a_n$ acquired from Planck CMB maps (see also Ref.~\cite{Buchert:2017uup} for a similar analysis). The coefficients were obtained by counting the occurrences of $\Theta$ values in Planck's SMICA temperature map. Here we chose two possible values for $\sigma_\Theta^2$: the sample variance computed from the CMB map $\sigma_\Theta^2 = 1.50 \times 10^{-9}$, with which $a_2 = 0$, and the one preferred by simulations $\sigma_\Theta^2 = 1.62 \times 10^{-9}$. The grey contours show the intrinsic noise $\sigma(a_n)$ (1- and 2-$\sigma$ regions) resulting from 500 Gaussian simulations using CAMB~\cite{CAMB-web} with the cosmological parameters reported by Planck~\cite{Ade:2015xua} (with a beam resolution of $5$ arcmin FWHM), and $\sigma_\Theta^2 = 1.62 \times 10^{-9}$, which is the average over simulations of the sample variances. As one might have expected, the observed values are mostly compatible with a Gaussian distribution. To get the $b_n$ coefficients via Eq.~(\ref{coeff-b_n}), we set $\Delta N = 60$ and fix $\ln \xi = 8$, which corresponds to the range of momenta $10^{-4}\;{\rm Mpc}^{-1} \leq k \leq 0.3\; {\rm Mpc}^{-1}$ for the observed modes in the CMB~\cite{Ade:2015lrj, Ade:2015hxq}. Given that Fig.~\ref{fig:a_n-Planck} lacks a conclusive imprint of non-Gaussianity, the potential in Fig.~\ref{fig:DeltaV-reconstructed} serves for illustrative purposes only. However, we must note that this type of analysis is cosmic variance limited, as evidenced by the different results obtained from the two values chosen for $\sigma_\Theta^2$. Additionally, there are a number of anomalies present in the CMB that we disregard herein, such as the statistical differences between the north and south hemispheres~\cite{Ade:2015hxq}. Nevertheless, we encourage the community to keep an eye out for these signatures, as well as to perform more sophisticated analyses with available data sets. For instance, one approach to try and circumvent the aforementioned effects is to compute the transfer functions for a restricted multipole range, which can be done by modifying accordingly the sums in Eqs.~\eqref{G-T} and~\eqref{F-T}, then to consider a filtered CMB map that only contains those contributions, and finally use Eq.~\eqref{coeff-b_n} as before to obtain the reconstructed potential.

The NG studied here has a fixed shape of the local type [recall Eq.~(\ref{n-point-local})], meaning that any relevant information is entirely contained in the coefficients $h_n$, related to the $a_n$'s via~(\ref{a-n-h-n}). However, the zero-lag cumulants approach offered in this Letter might not constitute the most efficient strategy to constrain the $h_n$'s. Shapes other than local, present in the data, will contribute to the measured $a_n$ cumulants, increasing the uncertainty on the deduced values of the $h_n$'s. Hence, to break the shape degeneracy hidden in the cumulants, more sophisticated techniques may be considered. For instance, following similar steps to those described here, one could derive the full probability functional containing information about the local shape (or other shapes, in the case of nontrivial interactions not considered here) to perform reconstructions.

Our methods may be repeated to attempt reconstructions employing LSS, 21 cm, and CMB spectral distortion data. The main difference would rest on the treatment of specific transfer functions needed to connect the $h_n$'s with new $a_n$ cumulants parametrizing new types of distributions (e.g., matter distribution in the case of LSS surveys). Granted that foregrounds and secondary NG's can be accurately modeled, these surveys should offer us the opportunity to perform better reconstructions of the landscape potential for the same reasons that they will improve upon current CMB constraints on the $f_{\rm NL}^{\rm local}$ parameter [i.e., reducing the uncertainty $\sigma(f_{\rm NL}^{\rm local})$]: they will give us access to a broader range of scales and/or larger data sets, allowing us to perform statistics with sharper cumulant uncertainties $\sigma(a_n)$. In this respect, it is worth recalling that soon to come LSS surveys will be able to reduce $\sigma(f_{\rm NL}^{\rm local})$ by a factor of $5-10$~\cite{Carbone:2010sb, Dore:2014cca, Raccanelli:2015oma}, whereas future 21 cm and CMB spectral distortion experiments promise to do so by factors $\sim 10^2$~\cite{Loeb:2003ya, Munoz:2015eqa, Meerburg:2016zdz} and $\sim 10^3$~\cite{Pajer:2012vz}, respectively. An important pending challenge is to understand to what degree a reduction of $\sigma(f_{\rm NL}^{\rm local})$ will come together with a reduction of the $\sigma(a_n)$'s.

To summarize, we have analyzed a novel class of primordial signatures that deserves to be thoroughly studied both theoretically and observationally, particularly on the wake of new CMB~\cite{Abazajian:2016yjj} and LSS~\cite{Euclid-web, Lsst-web} surveys.  Multifield models of inflation allow for regimes in which the statistics of isocurvature fields are transferred to $\zeta$, encoding information about the shape of the inflationary landscape potential in the observable curvature perturbations. We have considered a sufficiently generic situation described by the Lagrangian~(\ref{Lagrantian-full-v}), however, the transfer mechanism might be even more generic, and as such, constraining this type of NG has an enormous potential for characterization of the early Universe.

\begin{acknowledgments}

\emph{Acknowledgments}: We wish to thank Basti\'an Pradenas, Walter Riquelme and Domenico Sapone for useful discussions and comments. GAP and BSH acknowledge support from the Fondecyt Regular Project No.~1171811 (CONICYT). BSH is supported by a CONICYT Grant No.~CONICYT-PFCHA/Mag\'{i}sterNacional/2018-22181513. SS is supported by the Fondecyt Postdoctorado Project No.~3160299 (CONICYT).

\end{acknowledgments}

\end{document}